\documentclass[11pt,a4paper]{article}

\usepackage{jheppub}
\usepackage[T1]{fontenc}
\usepackage{lmodern}
\usepackage[utf8]{inputenc}
\usepackage{mathtools}
\usepackage{xcolor}
\usepackage[hang,flushmargin]{footmisc}

\newcommand{\setmatrixspacing}[2]{%
  \AddToHook{env/pmatrix/begin}{\setlength{\arraycolsep}{#1}\renewcommand{\arraystretch}{#2}}}
\setmatrixspacing{5pt}{1.0}

\newcommand{\kcs}{k_{\rm CS}}
\newcommand{\sg}{\mathrm{SG}}

\newcommand{\Tr}{\operatorname{Tr}}
\newcommand{\dn}{\operatorname{dn}}
\newcommand{\sn}{\operatorname{sn}}
\newcommand{\am}{\operatorname{am}}

\title{Black hole entropy from angular holonomies of Sine--Gordon soliton lattices in AdS$_3$}

\author[a]{Marcela C\'ardenas}
\author[b]{Francisco Correa}
\author[b]{Nicol\'as Salazar}

\affiliation[a]{Universidad San Sebasti\'an, Facultad de Ingenier\'ia, Bellavista 7, Recoleta, Santiago, Chile}
\affiliation[b]{Departamento de F\'isica, Universidad de Santiago de Chile, Avenida Victor Jara 3493, Estaci\'on Central, 9170124, Santiago, Chile}

\emailAdd{marcela.cardenasl@uss.cl}
\emailAdd{francisco.correa.s@usach.cl}
\emailAdd{nicolas.salazar.v@usach.cl}

\abstract{We construct boundary conditions for AdS$_3$ gravity that realize Sine–Gordon dynamics. For a periodic soliton lattice at fixed winding, we relate the angular holonomy to the exact acoustic dispersion by continuing the Bloch momentum through the zero-frequency band edge associated with rigid translation of the lattice. For suitable embedding angles, the stable vacuum has a regular BTZ geometry. When only one chiral sector carries winding, we construct a family whose asymptotic region extends to a BTZ black hole. As the spectral parameter of the winding sector tends to the corresponding band edge, the black hole approaches extremality. The leading correction to the extremal entropy is linear in the Hawking temperature defined with respect to laboratory time, with a coefficient inversely proportional to the sound velocity of the lattice. The same inverse-velocity dependence appears in the low-temperature entropy of one-dimensional quantum fluids.}
 

\notoc

\begin{document}
\maketitle
\flushbottom

\section{Introduction}

Three-dimensional Einstein gravity has no local propagating gravitational degrees of freedom, yet its boundary conditions can support nontrivial dynamics.  Since the Brown--Henneaux analysis, the physical content of asymptotically AdS$_3$ gravity has been known to depend crucially on boundary conditions and their asymptotic symmetries \cite{BrownHenneaux,CoussaertHenneauxvanDriel}.  In the Chern--Simons formulation \cite{AchucarroTownsend,Witten1988}, this freedom is encoded in the boundary behavior of two flat $SL(2,\mathbb R)$ connections \cite{Banados1995}.  A broad range of alternative AdS$_3$ boundary conditions and enhanced asymptotic symmetry structures has since been developed \cite{HenneauxPerezTempoTroncoso2013,BunsterEtAl2014,CompereSongStrominger,Troessaert2013,GrumillerRiegler2016,GonzalezMatulichPinoTroncoso2018}, including KdV-type and near-horizon/soft-hair conditions \cite{PerezTempoTroncoso2016,DonnayGiribetGonzalezPino2016,AfsharEtAl2016,GrumillerEtAl2020,GrumillerMerbis2020}.  Integrable hierarchies provide a particularly broad extension of this idea, with several nonlinear flows realized as gravitational boundary dynamics \cite{OjedaPerez2019,OjedaPerez2020,FuentealbaEtAl2018,LaraPinoReyes2024,PinoReyes2025,AdamiLatifi2025,AdamiLaraLatifiMeyer2026}. Applications include black-hole microstates and KdV-charged sectors \cite{MelnikovEtAl2019,DymarskySugishita2020}, as well as conductance and memory effects \cite{CardenasGonzalezLaraPino2022}. Related work addresses the integrability of gravitational charges \cite{RuzziconiZwikel2021} and the Weyl geometry of the boundary \cite{ArenasHenriquezEtAl2025}.  Our starting point is the class of boundary conditions introduced in Ref. \cite{IntegrableSpacetime}, in which Einstein's equations reduce to two copies of nonlinear integrable dynamics.  Their black-hole sector and thermodynamics were subsequently developed in Ref. \cite{IntegrableBH}, where stationary temperatures were shown to be encoded directly in spectral data.

Here we ask whether the speed of a disturbance travelling along a periodic soliton lattice can be read from the angular holonomy of the gravitational connection. Such lattices arise as stationary solutions of the Sine--Gordon equation with winding: as one goes around the spatial circle, the field winds through a whole number of periods. Both these configurations and their small fluctuations can be studied using the spectral theory of the periodic Sine--Gordon equation \cite{FaddeevTakhtajan,McKean,ForestMcLaughlin,ErcolaniForestMcLaughlin}. An earlier gravitational application related Sine--Gordon solitons to black holes in two-dimensional Jackiw--Teitelboim gravity \cite{GegenbergKunstatter}. The model also appears in a different AdS$_3$ setting through the relation between massless string and Sine--Gordon scattering matrices \cite{Torrielli2026}.

Translating the entire soliton lattice gives another stationary solution. When the displacement instead varies slowly from cell to cell, it travels along the lattice as a sound wave. Its speed can be obtained either from the compressibility and inertia of the lattice or from the slope of its lowest fluctuation band \cite{AristovLuther2002}. In a one-dimensional quantum fluid with a single sound mode, a slower wave means more modes in a given low-frequency interval and hence a larger low-temperature entropy density \cite{ChenEtAl2019}.

In each Chern--Simons sector, parallel transport around the spatial circle defines an angular holonomy whose eigenvalues give the chiral holonomy entropy. To see how the lattice enters this observable, consider the phase a sound wave accumulates over one cell. Continued through the band edge, this phase gives the spatial growth exponent that determines the holonomy. We compare embeddings of the same lattice, each at a fixed spectral parameter. Near the edge, a slower sound wave gives a steeper rise in entropy with the scale set by the temporal connection, as shown in Eq. \eqref{eq:entropy-vs}. In the mixed winding--vacuum sector, a regular Lorentzian collar admits a BTZ extension, and this relation becomes a horizon-entropy slope with respect to laboratory temperature.

For the simplest gravitational Sine--Gordon embedding, the stable constant reference configuration has a degenerate dreibein.  We show that two fixed embedding parameters repair this problem without changing the Sine--Gordon equations or the finite-gap data of either integrable copy.  For the causal choice of embedding parameters specified below, the stable constant configuration is represented by a regular hyperbolic BTZ geometry.  The modification is dynamically trivial for each separate auxiliary problem but geometrically nontrivial when the two Chern--Simons sectors are assembled into a metric.

Section~\ref{sec:embedding} presents the Sine--Gordon connections and the regular BTZ reference vacuum. Section~\ref{sec:winding} describes the stationary winding sectors. Section~\ref{sec:canonical} defines the laboratory energy and the holonomy entropy. Sections~\ref{sec:velocity} and \ref{sec:thermal} derive the acoustic dispersion and relate it to the Floquet entropy. Appendix~\ref{app:btz} treats the vacuum and mixed-sector BTZ geometries. Appendix~\ref{app:duality} derives the spectral duality and the sign of the lower-branch multipliers.

\section{Sine--Gordon dynamics and its gravitational embedding}
\label{sec:embedding}

In this section we introduce the Chern--Simons description of AdS$_3$ gravity and choose boundary conditions under which the flatness equations of each copy become the Sine--Gordon equation. In its simplest form this embedding assigns a degenerate metric to the stable Sine--Gordon vacuum. A constant rotation in each copy removes the degeneracy without changing the integrable dynamics. The vacuum then becomes a regular BTZ black hole.

\subsection{Chern--Simons formulation and conventions}
\label{sec:cs-setup}

In the Chern--Simons formulation \cite{AchucarroTownsend,Witten1988}, three-dimensional Einstein gravity with negative cosmological constant is described by two $sl(2,\mathbb R)$ connections $A^\pm$, with action
\begin{equation}
 I=I_{\rm CS}[A^+]-I_{\rm CS}[A^-],\qquad
 I_{\rm CS}[A]=\frac{\kcs}{4\pi}\int\Tr\!\Big(A\wedge dA+\tfrac23A\wedge A\wedge A\Big),\qquad
 \kcs=\frac{\ell}{4G}.
 \label{eq:cs-action}
\end{equation}
Here $\ell$ is the AdS radius, $G$ the three-dimensional Newton constant and $\kcs$ the Chern--Simons level. The field equations state that both connections are flat, $F^\pm=dA^\pm+A^\pm\wedge A^\pm=0$.

We use the real two-dimensional representation
\begin{equation}
 L_0=\frac12\begin{pmatrix}-1&0\\0&1\end{pmatrix},
 \qquad
 L_1=\begin{pmatrix}0&-1\\0&0\end{pmatrix},
 \qquad
 L_{-1}=\begin{pmatrix}0&0\\1&0\end{pmatrix},
\label{eq:matrix-basis}
\end{equation}
which satisfies $[L_i,L_j]=(i-j)L_{i+j}$, with invariant form $\langle X,Y\rangle=\Tr(XY)$. The combination $L_1+L_{-1}$ generates the compact $SO(2)$ subgroup and will appear repeatedly below.

Spacetime has coordinates $(t,\rho,\phi)$, with $\phi\sim\phi+2\pi$ and $\rho$ a radial coordinate. We work in radial gauge \cite{CoussaertHenneauxvanDriel},
\begin{equation}
 A^+=b^{-1}(d+a^+)b,
 \quad
 A^-=b(d+a^-)b^{-1},
 \quad
 b=\operatorname{diag}\!\left((\rho/\ell)^{1/2},(\rho/\ell)^{-1/2}\right)=e^{-\log(\rho/\ell)L_0}.
 \label{eq:radial-gauge}
\end{equation}
The boundary connections $a^\pm=a_t^\pm dt+a_\phi^\pm d\phi$ depend only on $(t,\phi)$. The bulk connections are flat if and only if
\begin{equation}
 F_{t\phi}^\pm\equiv\partial_t a_\phi^\pm-\partial_\phi a_t^\pm+[a_t^\pm,a_\phi^\pm]=0.
 \label{eq:boundary-flatness}
\end{equation}
Boundary conditions specify how $a_t^\pm$ and $a_\phi^\pm$ depend on a set of boundary fields. Equation \eqref{eq:boundary-flatness} then becomes the equations of motion of those fields. Refs. \cite{IntegrableSpacetime,IntegrableBH} used this freedom to realize AKNS dynamics. In Sec.~\ref{sec:sg-lax} we use it to realize Sine--Gordon.

The dreibein and metric are
\begin{equation}
 e=\frac{\ell}{2}(A^+-A^-),
 \qquad
 g_{\mu\nu}=2\Tr(e_\mu e_\nu).
 \label{eq:dreibein-metric}
\end{equation}
A metric interpretation requires an invertible dreibein. On that sector, Chern--Simons gauge transformations act locally as diffeomorphisms combined with local Lorentz transformations \cite{Witten1988,Matschull}. Where the dreibein degenerates this correspondence fails, and gauge-equivalent connections need not describe equivalent geometries. Section~\ref{sec:embedding-angles} uses this point.

At fixed $t$, let $U^\pm(\phi)$ be the parallel-transport matrix defined by
\begin{equation}
 \partial_\phi U^\pm(\phi)=-a_\phi^\pm U^\pm(\phi),
 \qquad U^\pm(0)=\mathbf 1.
 \label{eq:transport}
\end{equation}
The angular holonomy is the transport matrix after one full turn,
\begin{equation}
 H_\phi^\pm\equiv U^\pm(2\pi)
 =\mathcal P\exp\!\left[-\int_0^{2\pi}a_\phi^\pm\,d\phi\right],
 \label{eq:holonomy-def}
\end{equation}
where $\mathcal P$ denotes path ordering. Because the radial dressing in Eq. \eqref{eq:radial-gauge} is $\phi$-independent, this is also the holonomy of the bulk connection up to conjugation. A constant conjugation $a^\pm\to Ra^\pm R^{-1}$ maps $H^\pm_\phi$ to $RH^\pm_\phi R^{-1}$. Its eigenvalues, and hence its conjugacy class, are therefore unchanged. When $|\Tr H_\phi^\pm|>2$ the holonomy is hyperbolic: it has two distinct real eigenvalues whose product is one. This is the class relevant to BTZ black holes \cite{BHTZ}.

\subsection{The Sine--Gordon Lax pair as a boundary connection}
\label{sec:sg-lax}

Let $u^+(t,\phi)$ and $u^-(t,\phi)$ be real fields on the boundary cylinder, one for each copy. We seek boundary conditions whose flatness equations are the Sine--Gordon equations
\begin{equation}
 u_{tt}^\pm-u_{\phi\phi}^\pm=\pm\sin u^\pm,
\label{eq:sg-lab}
\end{equation}
with the Sine--Gordon mass set to one. The two signs are related by the shift $u\to u+\pi$, so both copies carry the same Sine--Gordon dynamics. The standard auxiliary problem in laboratory coordinates \cite{AKNS1973,KaupLaboratory1975,DeconinckMcGillSegal} depends on the light-cone derivative

\begin{equation}
 q=\partial_t u+\partial_\phi u.
 \label{eq:q-def}
\end{equation}
In each copy, $q^s=(\partial_t+\partial_\phi)u^s$. We use $s=\pm1$ for the two chiral Chern--Simons copies and suppress the sector label within a single copy. For a fixed spectral parameter $\nu>0$, the connections are

\begin{align}
 a_\phi&=\frac12
 \begin{pmatrix}
 s\nu-\dfrac{\cos u}{4\nu} & -\dfrac q2+\dfrac{\sin u}{4\nu}\\[2mm]
 \dfrac q2+\dfrac{\sin u}{4\nu} & -s\nu+\dfrac{\cos u}{4\nu}
 \end{pmatrix},
 \label{eq:aphi-nu}\\[1mm]
 a_t&=\frac12
 \begin{pmatrix}
 s\nu+\dfrac{\cos u}{4\nu} & -\dfrac q2-\dfrac{\sin u}{4\nu}\\[2mm]
 \dfrac q2-\dfrac{\sin u}{4\nu} & -s\nu-\dfrac{\cos u}{4\nu}
 \end{pmatrix}.
 \label{eq:at-nu}
\end{align}

A direct computation gives
\begin{equation}
 F_{t\phi}=\frac14\big(u_{tt}-u_{\phi\phi}-s\sin u\big)\,(L_1+L_{-1}),
 \label{eq:flatness-sg}
\end{equation}
so flatness is equivalent to the Sine--Gordon equation \eqref{eq:sg-lab} for every value of $\nu$. Equations \eqref{eq:aphi-nu} and \eqref{eq:at-nu} are the Sine--Gordon Lax pair, and $\nu$ is its spectral parameter.

\subsection{Embedding angles}
\label{sec:embedding-angles}
The unrotated Lax connections assign a degenerate dreibein to the stable constant vacuum. We now introduce the embedding angles used to obtain a regular BTZ representative.

Flatness and the holonomy spectrum are both invariant under constant conjugation. Any constant rotation of the Lax pair therefore realizes the same Sine--Gordon dynamics. In each copy we introduce an angle $\theta$ and the rotation
\begin{equation}
 R_s(\theta)=\exp\!\Big[\frac{s\theta}{2}(L_1+L_{-1})\Big]
 =\begin{pmatrix}\cos(\theta/2)&-s\sin(\theta/2)\\[1mm] s\sin(\theta/2)&\cos(\theta/2)\end{pmatrix}
 \in SO(2)\subset SL(2,\mathbb R).
\end{equation}
It is convenient to trade $(\nu,\theta)$ for the Cartesian parameters
\begin{equation}
 \lambda=\nu\cos\theta,\qquad c=\nu\sin\theta,\qquad \nu^2=\lambda^2+c^2.
 \label{eq:lambda-c}
\end{equation}
As a result, the rotated connection is given by
\begin{equation}
 a_\mu(\lambda,c;u)\equiv R_s(\theta)\,a_\mu(\nu,0;u)\,R_s(\theta)^{-1},
\label{eq:constantconjugation}
\end{equation}
where $a_\mu(\nu,0;u)$ denotes Eqs. \eqref{eq:aphi-nu}--\eqref{eq:at-nu}, takes the explicit form
\begin{equation}
a_\phi=\frac12
\begin{pmatrix}
s\lambda-\alpha &
 c-\frac12q-\alpha_u\\
 c+\frac12q-\alpha_u &
-s\lambda+\alpha
\end{pmatrix},
\qquad
a_t=\frac12
\begin{pmatrix}
s\lambda+\alpha &
 c-\frac12q+\alpha_u\\
 c+\frac12q+\alpha_u &
-s\lambda-\alpha
\end{pmatrix},
\label{eq:aphi}
\end{equation}
with
\begin{equation}
 \alpha(u)=\frac{\lambda\cos u+s\,c\sin u}{4\nu^2}=\frac{\cos(u-s\theta)}{4\nu},
 \qquad
 \alpha_u=\frac{d\alpha}{du}.
 \label{eq:alpha}
\end{equation}
At fixed \(\nu\), constant conjugation preserves the spectral data of each copy. We evaluate them in the unrotated representative \(\theta=0\), where \(c=0\) and \(\lambda=\nu\). The two independent bulk transformations generally change the dreibein and the metric. A common Lorentz rotation would preserve the metric. When both copies appear we restore the sector labels.

In bulk Chern--Simons terms, the two families are related by the gauge transformations $b^{-1}R_+b$ and $bR_-b^{-1}$. The lower-left entry of $b^{-1}R_+ b$ and the upper-right entry of $bR_- b^{-1}$ grow linearly with $\rho$ for nontrivial rotations. These bulk transformations can therefore change the asymptotic geometry. The unrotated vacuum illustrates this distinction explicitly.

\subsection{The stable vacuum is a BTZ black hole}
\label{sec:btz-vacuum}


To study the vacuum configuration and its stability, we take a minimum of the potential in each copy. With the signs of Eq. \eqref{eq:sg-lab} this means
\begin{equation}
 u_0^+=\pi,
 \qquad
 u_0^-=0.
\label{eq:sgvacuum}
\end{equation}
For $u^\pm=u_0^\pm+\varepsilon^\pm(t,\phi)$, linearization gives $\varepsilon^\pm_{tt}-\varepsilon^\pm_{\phi\phi}+\varepsilon^\pm=0$. A mode $\varepsilon^\pm\propto e^{-i\omega t+ip\phi}$, with $p\in\mathbb Z$, has $\omega^2=p^2+1$. All frequencies are real, so these vacua are linearly stable in the laboratory coordinates $(t,\phi)$. The opposite constant choices have an unstable homogeneous mode.

On the stable constant solution $q_\pm=0$, and Eq. \eqref{eq:aphi} gives
\begin{equation}
 a_\phi^\pm
 =\kappa_\pm
 \begin{pmatrix}
 \pm\cos\theta_\pm &\sin\theta_\pm\\ \sin\theta_\pm&\mp\cos\theta_\pm
 \end{pmatrix},
 \qquad
 \kappa_\pm=\frac12\left(\nu_\pm+\frac{1}{4\nu_\pm}\right)\geq\frac12,
 \label{eq:kappa}
\end{equation}
a constant traceless matrix with eigenvalues $\pm\kappa_\pm$. The temporal components are proportional to it,
\begin{equation}
 a_t^\pm=\chi_\pm a_\phi^\pm,
 \qquad
 \chi_\pm=\frac{4\nu_\pm^2-1}{4\nu_\pm^2+1},
\label{eq:chi-vac}
\end{equation}
so $a^\pm=a_\phi^\pm\,dx^\pm$ along the null coordinates
\begin{equation}
 x^\pm=\phi+\chi_\pm t=\Phi\pm\frac{T}{\ell},
 \qquad
 \Phi=\frac{x^++x^-}{2},
 \qquad
 \frac{T}{\ell}=\frac{x^+-x^-}{2}.
\label{eq:vac-coordinates}
\end{equation}
The map $(t,\phi)\mapsto(T,\Phi)$ has Jacobian $\ell(\chi_+-\chi_-)/2$. It is invertible when $\nu_+\neq\nu_-$, and at fixed $T$ it preserves the identification, $\Phi\sim\Phi+2\pi$.

In our basis the canonical BTZ connections are \cite{Banados1995}

\begin{equation}
 a^+_{\rm BTZ}=\big(L_{-1}-\kappa_+^2L_1\big)dx^+,
 \qquad
 a^-_{\rm BTZ}=\big(L_{1}-\kappa_-^2L_{-1}\big)dx^-.
 \label{eq:btz-connections}
\end{equation}
The vacuum connections are constant $SL(2,\mathbb R)$ conjugates of these,
\begin{equation}
 a^\pm=G_\pm\,a^\pm_{\rm BTZ}\,G_\pm^{-1},
 \qquad
 G_\pm=R_\pm(\theta_\pm)\,P_\pm^{-1},
 \qquad
 \label{eq:vac-to-btz}
\end{equation}
where
\begin{equation}
 P_+=\frac{1}{\sqrt{2\kappa_+}}\begin{pmatrix}\kappa_+&-\kappa_+\\1&1\end{pmatrix},
 \quad
 P_-=\frac{1}{\sqrt{2\kappa_-}}\begin{pmatrix}1&-1\\\kappa_-&\kappa_-\end{pmatrix}.
 \label{eq:Pplus-minus}
\end{equation}
The columns of each matrix are eigenvectors of the corresponding BTZ angular connection and $R_\pm(\theta_\pm)$ is the embedding rotation of Sec.~\ref{sec:embedding-angles}. The two families therefore have conjugate angular holonomies.

The BTZ parameters and Sine--Gordon parameters can be compared using Eqs. \eqref{eq:rpm} and \eqref{eq:vac-btz}, where mass and angular momentum are 
\begin{equation}
\ell M_{\rm BTZ}=\kcs(\kappa_+^2+\kappa_-^2),
\qquad
J_{\rm BTZ}=\kcs(\kappa_-^2-\kappa_+^2).
\end{equation}
The causal laboratory patch with $T$ increasing requires $c_+c_-<0$ and $\nu_-<1/2<\nu_+$, as derived in Appendix~\ref{app:btz}, Eq. \eqref{eq:physicalpatch}. The same parameters can be read off without any chart. Since $a_\phi^\pm$ is constant, $U^\pm(\phi)=\exp(-\phi\,a_\phi^\pm)$, and
\begin{equation}
 \operatorname{spec}(H_\phi^\pm)=\{e^{-2\pi\kappa_\pm},e^{2\pi\kappa_\pm}\},
 \qquad
 \Tr H_\phi^\pm=2\cosh(2\pi\kappa_\pm)>2.
\end{equation}
The holonomy exponents $\kappa_\pm$ are thus exactly the combinations $(r_+\mp r_-)/(2\ell)$. For the BTZ reference family, the pair of angular holonomies represents the angular identification in the $SL(2,\mathbb R)\times SL(2,\mathbb R)$ cover of the connected AdS$_3$ isometry group. Its conjugacy classes fix the quotient parameters \cite{BHTZ}. Appendix~\ref{app:btz} gives the radial domain of the laboratory chart and its relation to the BTZ horizon. For winding configurations the angular holonomies remain well defined, while the existence of a regular exterior requires a geometric check, carried out for the mixed family in Appendix~\ref{app:btz}.
Having identified the BTZ reference vacuum, we now turn to nonconstant Sine--Gordon configurations.

\section{Winding sectors and spectral backgrounds}
\label{sec:winding}

The Sine--Gordon field need not be strictly periodic on the angular circle.  It is enough that the physical connection and the metric be single-valued.  Since they depend on derivatives of $u$ and on $\sin u,\cos u$, the boundary condition
\begin{equation}
 u^\pm(t,\phi+2\pi)=u^\pm(t,\phi)+2\pi n_\pm,
 \qquad n_\pm\in\mathbb Z,
\label{eq:windingbc}
\end{equation}
leaves all local gravitational fields periodic.  The integers $n_\pm$ label disconnected topological sectors of the Sine--Gordon fields.

We focus on stationary configurations and the corresponding first integrals for each chiral sector, 
\begin{equation}
 u_t^\pm=0,
 \qquad
 u_{\phi\phi}^\pm=\mp\sin u^\pm, \qquad  \frac12(u_\phi^\pm)^2\mp\cos u^\pm=\delta^\pm.
 \label{eq:firstintegral}
\end{equation}
We consider the rotational elliptic solutions of the periodic Sine--Gordon equation \cite{DeconinckMcGillSegal}. Reversing the winding
in either sector leaves its fluctuation spectrum and angular monodromy eigenvalues unchanged. We first take a common positive winding
$n_+=n_-=n>0$, so the closure condition fixes the same elliptic modulus $k_n$ in both sectors. We also consider the mixed sector $(n_+,n_-)=(n,0)$, in which the $+$ copy is a winding lattice and the $-$ copy is the stable constant vacuum $u^-=0$, with $\delta^-=1$. For winding $n$, in our conventions one may choose, up to translations $\phi_\pm$,
\begin{equation}
 u_n^+(\phi)=2\,\am\!\left(\frac{\phi-\phi_+}{k_n},k_n\right),
 \qquad
 u_n^-(\phi)=\pi+2\,\am\!\left(\frac{\phi-\phi_-}{k_n},k_n\right),
\label{eq:winding-solution}
\end{equation}
with the angular derivatives 
\begin{equation}
 \partial_\phi u_n^\pm(\phi)
 =\frac{2}{k_n}\,
 \dn\!\left(\frac{\phi-\phi_\pm}{k_n},k_n\right).
 \label{eq:winding-derivative}
\end{equation}
For the common-winding profiles, imposing $n$ windings around the circle of circumference $2\pi$ and substituting the profiles into Eq. \eqref{eq:firstintegral} gives
\begin{equation}
 k_n K(k_n)=\frac{\pi}{n},
 \qquad
 \delta^+=\delta^-=\delta_n=\frac{2}{k_n^2}-1 .
 \label{eq:closure}
\end{equation}
The elliptic modulus is therefore a finite-volume density parameter.  Restoring a common Sine--Gordon mass $m$ on a circle of length $L$ gives $kK=mL/(2n)$, so the same family contains a controlled large-$mL$ regime.

\subsection{Boundary metric in winding sectors}
For the constant solution we checked explicitly that the two chiral connections define a Lorentzian metric near infinity. On the other hand, for a winding solution the same check must be made at each angle, since the connections now vary around the circle. In the radial gauge used here, conjugation by $b$ multiplies the lower-left entry of $a^+$ and the upper-right entry of $a^-$ by $\rho/\ell$ in the respective bulk connections. These entries therefore supply the leading tangential components of the dreibein. Denoting the corresponding boundary one-forms by $e^\pm$, the trace prescription gives the boundary line element
\begin{equation}
 e^+=(a^+_\mu)_{21}dx^\mu,\qquad
 e^-=(a^-_\mu)_{12}dx^\mu,\qquad
 ds_{(0)}^2\equiv\lim_{\rho\to\infty}\rho^{-2}ds^2\big|_{d\rho=0}
 =-e^+e^- ,
 \label{eq:winding-boundary-frame}
\end{equation}
where the product in the line element is symmetric. When the two forms are independent, their kernels define the two distinct null directions of a Lorentzian boundary metric. A direct test of independence is their area form,
\begin{equation}
 e^+\wedge e^- = D(\phi)\,dt\wedge d\phi,\qquad
 D=e_t^+e_\phi^- -e_\phi^+e_t^-,\qquad
 \det g^{(0)}=-\tfrac14 D^2 .
 \label{eq:winding-boundary-area}
\end{equation}
At a zero of $D$ the boundary metric loses rank. Multiplication by a smooth, nonzero conformal factor preserves this rank loss.

One can locate intervals containing such zeros for the aligned profiles of Eq. \eqref{eq:winding-solution}. When $\phi_+=\phi_-$, write $u=u_n^+$, so that $u_n^-=u+\pi$ and both profiles have the same positive derivative $q=\partial_\phi u$. Substitution of Eq. \eqref{eq:aphi} into Eq. \eqref{eq:winding-boundary-area} gives
\begin{equation}
 2D=\big(c_- -q/2\big)\alpha_u^+(u)
     -\big(c_+ +q/2\big)\alpha_u^-(u+\pi).
 \label{eq:winding-boundary-test}
\end{equation}
During each winding, $u$ increases by $2\pi$. At the successive phases $u=\pi/2$ and $u=3\pi/2$ (modulo $2\pi$), the first integral \eqref{eq:firstintegral} gives the same value $q=\sqrt{2\delta_n}$, while both trigonometric factors $\alpha_u^\pm$ change sign. Hence the values of $D$ at successive test angles are opposite. If they are nonzero, continuity gives a zero between each consecutive pair, including the last pair across the periodic identification. If they vanish, the test angles themselves are zeros. Thus, for $n_+=n_-=n>0$ and $\phi_+=\phi_-$, the boundary metric loses rank at no fewer than $2n$ distinct angles around the circle.

This argument assumes aligned profiles, $\phi_+=\phi_-$. A relative translation preserves the angular holonomy eigenvalues of each copy but changes $D(\phi)$, so the rank of the boundary metric must be checked separately for shifted profiles.

A different possibility arises in the mixed sector $(n,0)$. Taking $u^-=0$, $\theta_+=\pi/2$, $\theta_-=-\pi/2$, the upper winding branch in the $+$ copy and $0<\nu_-<1/2$ in the vacuum copy, with equal positive temporal eigenvalue magnitudes $\zeta$, gives a Lorentzian collar with spacelike angular circles and timelike $\partial_t$. Appendix~\ref{app:btz} constructs its extension to a BTZ exterior and its smooth Euclidean filling. This supplies a horizon interpretation of the holonomy entropy for this mixed family.

The entries of $a_\phi^\pm$ remain smooth even at the degeneracies of the aligned profiles. Parallel transport around the full circle is therefore well defined, and its eigenvalues still enter the holonomy entropy of Sec.~\ref{sec:entropy-scope}. This allows us to compute the charges and spectral data directly from the connections. A black-hole interpretation requires a further geometric check: the connections must admit a representative with an invertible dreibein throughout the Lorentzian exterior and a smooth Euclidean continuation in which the thermal circle contracts, as in the mixed-sector extension of Appendix~\ref{app:btz}.

\section{Gravitational energy and holonomy entropy}
\label{sec:canonical}

We first identify the gravitational charge that generates translations
in laboratory time $t$. We then turn to the entropy functional, which
is determined by parallel transport around the angular circle. The
energy calculation involves both components of the boundary connection:
$a_t^\pm$ specifies the time evolution, while $a_\phi^\pm$ is the field
whose variation enters the charge.

Consider one chiral sector and write $s=+1$ for the $+$ copy and
$s=-1$ for the $-$ copy. We suppress the sector label on the fields
until the final expression. Let $\Sigma$ be a constant-$t$ slice,
with outer boundary oriented along increasing $\phi$. We hold the embedding parameters $\lambda$ and $c$ fixed when varying the fields. Following the canonical Chern--Simons charge construction \cite{Banados1995}, we use $\delta Q_\epsilon=\Omega_s(\delta A,D\epsilon)$. With our orientation
and level conventions, the Chern--Simons presymplectic form in this
sector is
\begin{equation}
 \Omega_s(\delta_1 A,\delta_2 A)
 =\frac{s\kcs}{2\pi}\int_\Sigma
   \Tr\!\left(\delta_1 A\wedge\delta_2 A\right).
 \label{eq:cs-presymplectic-energy}
\end{equation}
Here $A$ denotes the bulk connection, pulled back to $\Sigma$ in the
integral. To find the generator of a transformation, we pair its
variation with an arbitrary variation $\delta A$ tangent to the space
of solutions. For a gauge transformation $\delta_\epsilon A=D\epsilon$,
integration by parts gives
\begin{equation}
 \Omega_s(\delta A,D\epsilon)
 =\frac{s\kcs}{2\pi}\int_\Sigma\Tr(\epsilon\,\delta F)
  -\frac{s\kcs}{2\pi}\int_{\partial\Sigma}\Tr(\epsilon\,\delta A),
 \label{eq:cs-generator-boundary}
\end{equation}
where $F=dA+A\wedge A$ and $D$ is the covariant derivative defined
by $A$. All solutions satisfy $F=0$, so their allowed variations
satisfy $\delta F=0$. The generator therefore receives only boundary
contributions.

To identify the charge associated with time translations, consider a
spacetime vector field $\xi=\xi^\nu\partial_\nu$. The Lie derivative
of the connection along $\xi$ can be written in components as
\begin{equation}
 (\mathcal L_\xi A)_\mu
 =D_\mu(\xi^\nu A_\nu)+\xi^\nu F_{\nu\mu},
 \qquad
 D_\mu X=\partial_\mu X+[A_\mu,X].
 \label{eq:cs-diffeomorphism-energy}
\end{equation}
For a flat connection, $F_{\nu\mu}=0$, so the right-hand side reduces
to a gauge transformation with parameter $\epsilon=\xi^\nu A_\nu$.
For a translation in laboratory time, $\xi=\partial_t$, this parameter
is simply $\epsilon=A_t$. On the
outer boundary, the pullback of $\delta A$ to the angular circle is
$\delta A_\phi\,d\phi$. The radial conjugations of $A_t$ and
$\delta A_\phi$ cancel inside the trace, so
Eq. \eqref{eq:cs-generator-boundary} gives the variation of the
asymptotic time-translation charge directly in terms of the boundary
fields:
\begin{equation}
 \delta Q_t^{(s)}
 =-\frac{s\kcs}{2\pi}\oint d\phi\,
   \Tr\!\left(a_t\,\delta a_\phi\right).
 \label{eq:cs-laboratory-charge-variation}
\end{equation}
Although $a_t$ depends on the Sine--Gordon field, it enters here as
the parameter of the transformation evaluated on the given solution.
The variation $\delta a_\phi$ describes the displacement to a
neighboring solution.

We now evaluate this charge on the Sine--Gordon family, keeping the
embedding parameters $\lambda$ and $c$ fixed. Writing
$q=u_t+u_\phi$ and using Eq. \eqref{eq:aphi},
the matrix trace reduces to
\begin{equation}
 \Tr(a_t\,\delta a_\phi)
 =-\frac18 q\,\delta q
  -\frac12\bigl(s\lambda\alpha_u-c\alpha\bigr)\delta u
 =-\frac1{16}\,\delta\!\left(q^2+2s\cos u\right).
 \label{eq:cs-laboratory-charge-trace}
\end{equation}
In the first equality we used $\alpha_{uu}=-\alpha$. In the second
we used $s\lambda\alpha_u-c\alpha=-s\sin u/4$, which follows from
the definition of $\alpha$. The cancellation of the explicit embedding
parameters leaves an exact variation. Integrating in field space and
restoring the chiral labels gives
\begin{equation}
 Q_t^\pm
 =\pm\frac{\kcs}{32\pi}F^{(\pm)}+E_0^\pm(\nu_\pm),
 \qquad
 F^{(\pm)}
 =\oint d\phi\,\bigl(q_\pm^2\pm2\cos u^\pm\bigr).
 \label{eq:grav-hamiltonian}
\end{equation}
The additive constant $E_0^\pm$ fixes the energy assigned to the
chosen reference configuration in each sector.

The physical content of this result becomes transparent upon expanding
$q^2=(u_t+u_\phi)^2$. In the Sine--Gordon conventions used here,
the energy and momentum of one copy are
\begin{equation}
 H_{\sg}^{(s)}
 =\oint d\phi\,
  \left(\frac12 u_t^2+\frac12 u_\phi^2+s\cos u\right),
 \qquad
 P_{\sg}^{(s)}
 =\oint d\phi\,u_tu_\phi .
 \label{eq:sg-energy-momentum-laboratory}
\end{equation}
The cross term in $q^2$ supplies the momentum contribution, so that
\begin{equation}
 F^{(s)}=2\bigl(H_{\sg}^{(s)}+P_{\sg}^{(s)}\bigr).
 \label{eq:sg-lightcone-energy}
\end{equation}
Thus the gravitational charge generating laboratory-time translations
selects the light-cone combination of Sine--Gordon energy and
momentum in each sector. Its spacetime interpretation depends on the
choice of time coordinate. On the constant BTZ background,
Eq. \eqref{eq:vac-coordinates} expresses $\partial_t$ as a linear
combination of canonical BTZ time translation and angular rotation.

\subsection{Holonomy entropy}
\label{sec:entropy-scope}

The angular holonomy $H_\phi^\pm$ defined in Eq. \eqref{eq:holonomy-def}
is the monodromy matrix of the angular transport equation
\eqref{eq:transport}. We write $M^\pm\equiv H_\phi^\pm$ when discussing
its Floquet spectrum. To see the meaning of a Floquet multiplier,
choose an eigenvector $v_0$ of $M^\pm$ with eigenvalue $\Lambda$ and
transport it along the circle. Since the angular connection is
$2\pi$-periodic, the resulting solution satisfies
\begin{equation}
 v(\phi)=U^\pm(\phi)v_0,
 \qquad
 v(\phi+2\pi)=\Lambda\,v(\phi).
 \label{eq:angular-floquet-multiplier}
\end{equation}
Thus $|\Lambda|$ gives the factor by which the amplitude of this
solution changes after one full turn.

We consider the sector in which both angular holonomies are
hyperbolic,
\begin{equation}
 \bigl|\Tr M^\pm\bigr|>2.
 \label{eq:angular-hyperbolicity}
\end{equation}
In each chiral copy the two multipliers are real and reciprocal.
One solution therefore grows with each turn around the circle,
while the other decays. The growth factor per turn is the spectral
radius of the corresponding monodromy matrix,
\begin{equation}
 r(M)=\max\bigl\{|\Lambda_1|,|\Lambda_2|\bigr\}.
 \label{eq:angular-spectral-radius}
\end{equation}

The gravitational entropy functional \cite{deBoer:2013gz,Ammon:2013hba} sums the logarithmic growth
per turn in the two chiral sectors,
\begin{equation}
 S\equiv\kcs\bigl(\log r(M^+)+\log r(M^-)\bigr).
 \label{eq:entropy-holonomy-definition}
\end{equation}
Multiplying either monodromy by the central element $-\mathbf 1$
reverses the signs of its multipliers and leaves $S$ unchanged.

For Ba\~nados geometries with Brown--Henneaux boundary conditions, the two Floquet indices likewise enter the horizon entropy \cite{SheikhYavartanoo}. Here we use the holonomy functional on the Sine--Gordon winding family. Its normalization follows from the Euclidean gravitational entropy
\cite{BunsterEtAl2014,IntegrableBH,AfsharEtAl2016,GrumillerMerbis2020}.
For the constant vacuum, \(r(M^\pm)=e^{2\pi\kappa_\pm}\). Hence \(S=2\pi\kcs(\kappa_++\kappa_-)=2\pi r_+/(4G)\), so the sum of the two logarithmic holonomy exponents, weighted by the Chern--Simons level, gives the outer horizon circumference in units of $4G$. This motivates evaluating the same observable on winding solutions, where the connections vary around the circle and their monodromies can be computed exactly. For a winding background, a horizon interpretation also requires an invertible exterior and a compatible smooth Euclidean filling. In higher-spin gravity, thermal holonomy conditions characterize black-hole solutions \cite{GutperleKraus2011}, and a higher-spin gauge transformation can change whether the metric displays a smooth horizon \cite{AmmonEtAl2011}. The mixed family realizes these conditions through the extension in Appendix~\ref{app:btz}.

\section{Winding backgrounds and their acoustic fluctuations}
\label{sec:velocity}

A stationary winding solution describes a periodic array of Sine--Gordon solitons. The winding $n$ fixes how many periods fit on the circle, while the elliptic modulus $k_n$ fixes their shape through Eq. \eqref{eq:closure}. We now ask how a small disturbance propagates through this background. This provides the physical velocity that will enter the entropy relation.

Work first with the $-$ copy and $m=1$; shifting $u$ by $\pi$ gives the same fluctuation operator for the $+$ copy. For positive winding, write
\begin{equation}
 u_0(\phi)=\pi+2\am(\phi/k,k),
 \qquad
 u(t,\phi)=u_0(\phi)+\varepsilon e^{-i\omega t}\psi(\phi).
\end{equation}
Here $\varepsilon$ is an infinitesimal amplitude, $\omega$ the frequency, and $\psi$ the spatial profile of the perturbation. Negative winding is related by reflection and has the same spectrum. Keeping terms linear in $\varepsilon$ yields $-\partial_\phi^2\psi+\cos u_0\,\psi=\omega^2\psi$, or
\begin{equation}
 -\psi_{zz}+2k^2\sn^2(z,k)\psi=k^2(\omega^2+1)\psi,
 \qquad z=\phi/k.
\label{eq:lame}
\end{equation}
This is the one-gap Lam\'e equation: a periodic Schr\"odinger problem with a single finite spectral gap \cite{JonesMarangellMillerPlaza2013}.

The lower edge of the first Lam\'e band corresponds to $\omega^2=0$
in Eq. \eqref{eq:lame}. Its physical origin is the freedom to translate
the winding profile: for any constant $\phi_0$, the shifted profile
$u_0(\phi-\phi_0)$ is a static solution with the same energy.
Differentiating with respect to $\phi_0$ therefore gives a
zero-frequency fluctuation proportional to $\partial_\phi u_0$.
Choosing the origin at the center of a cell and fixing the overall
normalization, we write this mode as
\begin{equation}
 \psi_0(\phi)=\dn(\phi/k,k),
 \qquad a=2kK(k),
 \label{eq:lame-translation-mode}
\end{equation}
where $a$ is its period in the laboratory coordinate $\phi$.

\begin{figure}[htbp]
 \centering
 \includegraphics[width=0.95\linewidth]{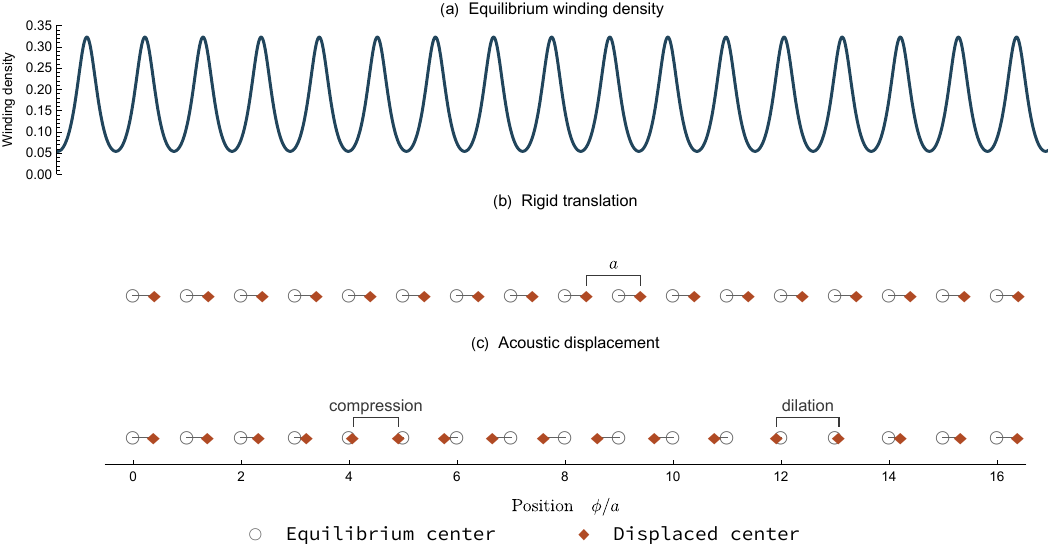}
 \caption{From rigid translation to an acoustic displacement of the
 Sine--Gordon soliton lattice. (a) The maxima of the equilibrium
 winding density $u'_0/(2\pi)$ locate the soliton centers.
 In (b) and (c), open circles mark the equilibrium positions and
 filled diamonds the displaced positions. (b) A rigid translation
 preserves the separation $a$ between neighboring centers.
 (c) On the extended lattice, a sinusoidal displacement of wavelength $16a$ produces the
 indicated regions of compression and dilation. The displacement
 amplitude is exaggerated for visibility.}
 \label{fig:soliton-displacements}
\end{figure}

This mode displaces all soliton centers by the same amount. To see how
a displacement propagates along the lattice, consider a fluctuation
with Bloch wave number $p$, defined by
\begin{equation}
 \psi_p(\phi+a)=e^{ipa}\psi_p(\phi).
 \label{eq:lame-bloch-condition}
\end{equation}
The sound velocity is the slope of the continuous Bloch band of the periodically extended lattice. On the circle of \(n\) cells, periodic perturbations sample the band at \(e^{ipna}=1\), or integer laboratory momenta modulo \(n\).
At $p=0$, all soliton centers undergo the same displacement and their
spacing remains equal to $a$. For $|p|a\ll1$, the displacement varies
slowly from cell to cell. As illustrated in
Fig.~\ref{fig:soliton-displacements}, neighboring centers move closer
together in one part of the lattice and farther apart in another.
This modulation of the soliton spacing is the acoustic mode.

The acoustic band of the semiclassical soliton lattice was obtained in Ref. \cite{AristovLuther2002}. The explicit Lam\'e solutions in Ref. \cite{CenCorreaFring2017} determine both the dispersion across the band and its slope at the translation mode. In their scalar Lam\'e equation the elliptic parameter is $k^2$ and the potential is shifted by $-(1+k^2)$ relative to Eq. \eqref{eq:lame}. Thus their eigenvalue is $k^2\omega^2-1$. With $\beta$ a complex spectral parameter, a reciprocal pair of Bloch solutions and its frequency are
\begin{equation}
 \psi_\pm(z;\beta)=\frac{H(z\pm\beta)}{\Theta(z)}
             \exp[\mp z Z(\beta,k)],\qquad
 \omega^2(\beta)=\frac{1}{k^2}-\sn^2(\beta,k).
 \label{eq:lame-bloch-pair}
\end{equation}
Here $k'=\sqrt{1-k^2}$, $K=K(k)$, $K'=K(k')$, and $H$ and $\Theta$ are the Jacobi theta functions $\vartheta_1$ and $\vartheta_4$ with argument $\pi z/(2K)$ and nome $\exp(-\pi K'/K)$. The Jacobi zeta function is $Z(z,k)=\partial_z\log\Theta(z)$. A translation by one cell takes $z$ to $z+2K$. Under this shift $H$ changes sign and $\Theta$ is periodic, so the cell multipliers are
\begin{equation}
 \mu_\pm(\beta)=-\exp[\mp 2K Z(\beta,k)],\qquad
 e^{ipa}=\mu_\pm(\beta).
 \label{eq:lame-cell-multipliers}
\end{equation}
Together, Eqs. \eqref{eq:lame-bloch-pair} and \eqref{eq:lame-cell-multipliers} give the dispersion parametrically. Unit-modulus multipliers describe propagating Bloch modes. The acoustic band runs from the periodic translation mode $\dn(z,k)$ at $\omega=0$ to the antiperiodic mode $\operatorname{cn}(z,k)$ at $\omega=k'/k$. The second allowed band starts at $\omega=1/k$ with the antiperiodic mode $\sn(z,k)$. 

The sound velocity measures how the frequency changes when neighboring
cells acquire a small relative phase. At the lower edge of the first
band, the fluctuation is the translation mode: every cell undergoes
the same displacement, so $\omega=0$ and the phase advance per cell
is $pa=0$. In the exact Bloch solution this mode occurs at
$\beta_0=K+iK'$. We approach it along the allowed band by writing
$\beta=\beta_0+iy$, with $y$ real and small, and compute $\omega$ and
$p$ along this same path.

For the frequency, the shift identity
$\sn(x+iK',k)=1/[k\sn(x,k)]$ gives
\begin{equation}
 \omega^2(\beta_0+iy)
 =\frac{1}{k^2}
  \left[1-\frac{1}{\sn^2(K+iy,k)}\right].
 \label{eq:lame-frequency-shift}
\end{equation}
Since $\sn(K+iy,k)=1+k'^2y^2/2+O(y^4)$, the squared frequency
vanishes quadratically in $y$.

At the lower band edge, the translation mode is periodic over one
cell, so $\mu_+(\beta_0)=1$. Its Bloch phase $pa$ is therefore zero
modulo $2\pi$. We choose the branch with $p(\beta_0)=0$ and write
the phase near the edge as
\begin{equation}
 i p a
 = \log\!\left[\frac{\mu_+(\beta)}{\mu_+(\beta_0)}\right]
 = -2K\bigl[Z(\beta,k)-Z(\beta_0,k)\bigr],
 \label{eq:lame-relative-phase}
\end{equation}
where the logarithm is continuous and vanishes at $\beta=\beta_0$.
Differentiating along $\beta=\beta_0+iy$ and using the Jacobi
identity $\partial_\beta Z(\beta,k)=\dn^2(\beta,k)-E/K$, with
$E=E(k)$, gives
\begin{equation}
 \frac{d(pa)}{dy}
 = -2K\,\partial_\beta Z(\beta,k)
 = 2E-2K\dn^2(\beta,k).
 \label{eq:lame-phase-derivative}
\end{equation}
The dispersion relation gives
$\dn^2(\beta,k)=1-k^2\sn^2(\beta,k)=k^2\omega^2(\beta)$.
Since $\omega^2$ vanishes quadratically in $y$, the phase derivative
is $2E+O(y^2)$. Integrating from $0$ to $y$ with $p(\beta_0)=0$
and combining the result with the frequency expansion, we obtain
\begin{equation}
 \omega^2=\frac{k'^2}{k^2}y^2+O(y^4),
 \qquad
 pa=2Ey+O(y^3).
 \label{eq:lame-edge-expansion}
\end{equation}

On the positive-frequency branch, $\omega=(k'/k)|y|+O(|y|^3)$.
Eliminating $y$ in favor of the physical wave number $p$ and using
the cell length $a=2kK$, we find the acoustic dispersion
\begin{equation}
 \omega(p)=v_s(k)|p|+O(|p|^3),
 \qquad
 v_s(k)=\frac{k'K(k)}{E(k)} .
 \label{eq:acoustic-velocity}
\end{equation}
For a lattice on the laboratory circle, the closure
condition \eqref{eq:closure} fixes $k=k_n$ and thus determines the
sound velocity at each winding. This agrees with the phonon velocity $k^\prime K/E$ of Aristov and Luther \cite{AristovLuther2002}, with soliton density $\bar\rho=n/L$ and $kK=m/(2\bar\rho)$.

\section{Holonomy entropy and Floquet spectral data}
\label{sec:thermal}

The dispersion relation of the winding lattice also determines its angular
holonomy. We first identify the frequency at which to evaluate the
Lam\'e solution of the preceding section.

Following the spectral construction of Ref.~\cite{IntegrableBH}, we extract the temporal eigenvalues from the determinant of the stationary connection. The matrix $a_t$ is
traceless, so its characteristic equation is
$\zeta^2=-\det a_t$. To evaluate the determinant, we may use the
$c=0$ representative with spectral parameter
$\nu=\sqrt{\lambda^2+c^2}>0$, since the constant conjugation relating
the two representatives preserves the determinant. For example, in
the $+$ copy, substitution of the stationary connection gives
\begin{equation}
 -\det a_t
 = \frac{\nu^2}{4}+\frac{1}{64\nu^2}
   -\frac18\left[\frac12(\partial_\phi u)^2-\cos u\right].
 \label{eq:temporal-det-first-integral}
\end{equation}
The expression in brackets is precisely the conserved first integral
$\delta_n$ in Eq. \eqref{eq:firstintegral}. This explains why the
temporal eigenvalues are independent of position even though the
winding profile varies around the circle. The other chiral copy gives
the same expression in terms of its own first integral $\delta^-$. For the stable vacuum $\delta^-=1$, whereas each nonzero winding uses its own $\delta_n$. To see what these eigenvalues measure, consider temporal transport
at a fixed angle $\phi$. In either chiral copy the auxiliary vector
$w$ obeys $\partial_t w=-a_t(\phi)w$. Since the background is
stationary, an eigenvector of $a_t(\phi)$ evolves exponentially
in laboratory time,
\begin{equation}
 a_t(\phi)w_\pm=\pm\zeta w_\pm,
 \qquad
 w_\pm(t,\phi)=e^{\mp\zeta t}w_\pm(0,\phi).
 \label{eq:auxiliary-temporal-evolution}
\end{equation}
The subscripts on $w_\pm$ label the two eigenvectors within a single
chiral copy. Since $a_t$ is traceless, its eigenvalues satisfy
\begin{equation}
 \zeta^2=-\det a_t=\frac{P_n(\nu)}{64\nu^2},
 \qquad
 P_n(\nu)=16\nu^4-8\delta_n\nu^2+1.
 \label{eq:temporal-spectral-polynomial}
\end{equation}
Here $P_n$ is the spectral polynomial of the stationary winding lattice. As in Ref.~\cite{IntegrableBH}, its value fixes the eigenvalue scale entering the Euclidean period. In the vacuum, $P_0(\nu)=(4\nu^2-1)^2$ gives $\zeta=|\chi|\kappa$, with a double root at $\nu=1/2$. On the winding branches where $P_n(\nu)>0$, we take $\zeta>0$. At fixed
winding, $\zeta$ vanishes at the band edge and provides a parameter
for the near-edge expansion of the entropy. The entropy itself is
determined by the multipliers of the angular monodromy
$M=U(2\pi)$, as in Eq. \eqref{eq:entropy-holonomy-definition}.

The subscript $\mathrm{ch}$ below denotes a single chiral copy. To express this expansion in thermal units, consider the continuation
$t=-i\tau$. If the chosen time flow generates a contractible
Euclidean cycle of dimensionless period $\beta_{\rm ch}$, the two
vectors in Eq. \eqref{eq:auxiliary-temporal-evolution} acquire the
factors $e^{\pm i\beta_{\rm ch}\zeta}$ after one circuit. These are
the eigenvalues of the holonomy around that cycle. On the standard
smooth branch the holonomy is $-\mathbf 1$, so its smallest positive
period satisfies $\beta_{\rm ch}\zeta=\pi$ \cite{IntegrableBH}.
This motivates the normalization
\begin{equation}
 \Theta_{\rm ch}\equiv\frac{\zeta}{\pi}
 =\frac{\sqrt{P_n(\nu)}}{8\pi\nu}.
 \label{eq:Tch}
\end{equation}
We use $\Theta_{\rm ch}$ to compare the angular entropy with the
temporal spectrum at fixed winding. A common elementary thermal cycle generated by $\partial_t$ requires $\zeta_+=\zeta_-$. When the two chiral connections
admit a compatible smooth thermal cycle generated by the chosen
flow, $\Theta_{\rm ch}=1/\beta_{\rm ch}$ in dimensionless
laboratory time \cite{IntegrableBH}. In physical time $\ell t$, the corresponding
inverse-period scale is $\Theta_{\rm ch}/\ell$.

The two temporal eigenvalues meet at $\zeta=0$. For a winding
background, the first integral has the value $\delta_n=2/k_n^2-1$,
so the condition $P_n(\nu)=0$ is a quadratic equation in $\nu^2$.
Its two positive roots are
\begin{equation}
 \nu_{\rm low}^{(n)}=\frac{1-k_n'}{2k_n},
 \qquad
 \nu_{\rm up}^{(n)}=\frac{1+k_n'}{2k_n},
 \qquad
 \nu_{\rm low}^{(n)}\nu_{\rm up}^{(n)}=\frac14,
 \qquad k_n'=\sqrt{1-k_n^2}.
 \label{eq:temporal-winding-edges}
\end{equation}
These values are the band edges in the auxiliary spectral
parameter $\nu$. Their connection to the physical fluctuation
spectrum will follow from the relation $\omega^2=-4\zeta^2$
derived below: both roots map to the zero-frequency translation
mode of the Lam\'e problem.

Between the two roots, $P_n(\nu)<0$ and the temporal eigenvalues
are imaginary. In this interval the angular Bloch multipliers over
one primitive cell have unit modulus, so a Bloch solution acquires
only a phase from cell to cell. The circle contains $n$ cells.
At isolated values of $\nu$, the phase accumulated over all cells
can therefore be an integer multiple of $\pi$, giving full-circle
monodromy $+{\bf 1}$ or $-{\bf 1}$. Appendix~\ref{app:duality}
discusses these resonances and the point $\nu=1/2$.

Outside the interval the angular multipliers are real and reciprocal,
so one solution grows and the other decays from cell to cell. The
causal patch of the BTZ reference vacuum, Eq. \eqref{eq:physicalpatch}, selects $\nu_+>1/2$ and $0<\nu_-<1/2$.
For common winding $n>0$, the ordering $\nu_{\rm low}^{(n)}<1/2<\nu_{\rm up}^{(n)}$ and hyperbolicity in both copies restrict this patch to
\begin{equation}
 \nu_+>\nu_{\rm up}^{(n)},
 \qquad
 0<\nu_-<\nu_{\rm low}^{(n)}.
 \label{eq:causal-hyperbolic-branches}
\end{equation}
At fixed winding, the edge values in $\nu$ depend only on $k_n$.
The embedding constant $c$ enters through
$\nu^2=\lambda^2+c^2$ and hence determines where a given edge is
reached as $\lambda$ is varied. At fixed $c$, this scan covers only $\nu\geq |c|$; an edge below $|c|$ requires a different choice of embedding.

\subsection{Floquet form of the holonomy entropy}
\label{sec:floquet-entropy}

Consider one copy on the upper hyperbolic branch. Stationary flatness
gives $a_t(\phi)=U(\phi)a_t(0)U(\phi)^{-1}$, so $M=U(2\pi)$
preserves the two real eigenlines of $a_t(0)$. An auxiliary solution
$w$ on either line satisfies
\begin{equation}
 \begin{aligned}
 a_t w&=\sigma\zeta w, & w(t,\phi)&=e^{-\sigma\zeta t}w(0,\phi),\\
 w(t,\phi+2\pi)&=\Lambda w(t,\phi), & \zeta&=\pi\Theta_{\rm ch},\quad \sigma=\pm1 .
 \end{aligned}
 \label{eq:zeta-theta}
\end{equation}
To read $\Lambda$ from the known Lam\'e solution, form
$f=w_1^2+w_2^2$ and $h=w_1^2-w_2^2$. In the $c=0$ representative
of the $+$ copy, adding and subtracting the two auxiliary equations
using Eqs. \eqref{eq:aphi-nu} and \eqref{eq:at-nu} gives
\begin{equation}
 (\partial_t+\partial_\phi)f=-2\nu h,\qquad
 (\partial_t-\partial_\phi)h=-\frac{\cos u}{2\nu}f .
\end{equation}
Thus $f_{tt}-f_{\phi\phi}=\cos u\,f$, precisely the fluctuation
equation solved in the preceding section. The shifted $-$ background
gives the same scalar operator. This is the squared-eigenfunction
relation for Sine--Gordon \cite{Kaup1984,DeconinckMcGillSegal}. Since
$f$ is quadratic in $w$, its time dependence is $e^{-2\sigma\zeta t}$
and its multiplier around the circle is $\Lambda^2$. The circle has
$n$ primitive cells. Hence the frequency and the one-cell Lam\'e
multiplier $\mu$ obey
\begin{equation}
 \omega^2=-4\zeta^2,\qquad \Lambda^2=\mu^n.
 \label{eq:spectral-map}
\end{equation}
The exact Lam\'e dispersion and multipliers in
Eqs. \eqref{eq:lame-bloch-pair} and \eqref{eq:lame-cell-multipliers}
give $\sn^2(\beta,k)=k^{-2}+4\zeta^2$ and
$\mu_\pm=-\exp[\mp2K Z(\beta,k)]$, where $k=k_n$.
Taking the larger modulus gives the exact chiral entropy
\begin{equation}
 S_{\rm ch}=\kcs\log r(M)
 =\kcs\, n\, K\,|\operatorname{Re}Z(\beta,k)| .
 \label{eq:holonomy-jacobi-zeta}
\end{equation}
For a real parametrization, set $\beta=x+iK'$ with $0<x<K$ and
$\sn(x,k)=(1+4k^2\zeta^2)^{-1/2}$. The imaginary half-period shift of
$\Theta$ gives $\operatorname{Re}Z(x+iK',k)=\partial_x\log H(x)$.
Using $H(x)/\Theta(x)\propto\sn(x,k)$, we obtain
\begin{equation}
 S_{\rm ch}=\kcs\,nK\left[Z(x,k)+
 \frac{\operatorname{cn}(x,k)\,\dn(x,k)}{\sn(x,k)}\right].
 \label{eq:holonomy-real}
\end{equation}
At the band edge $x\to K$, both terms vanish.

\begin{figure}[htbp]
 \centering
 \includegraphics[width=0.95\linewidth]{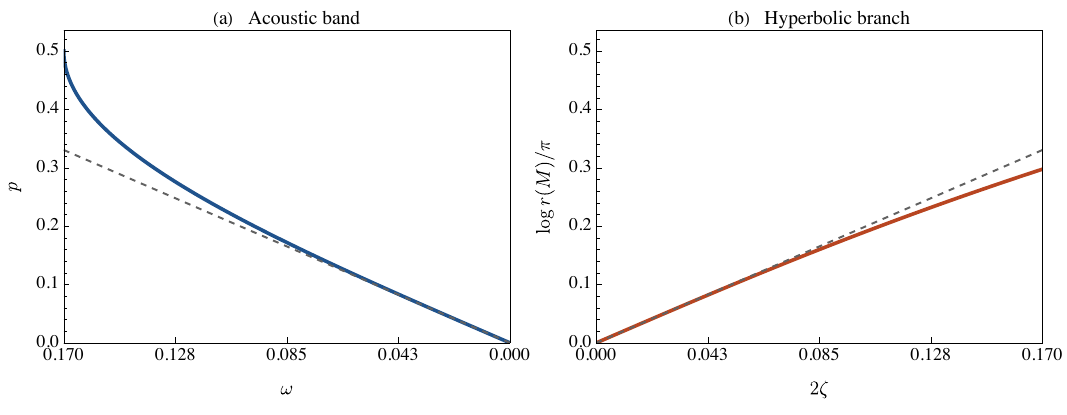}
 \caption{Acoustic dispersion and chiral holonomy entropy for
 winding $n=1$. The two panels meet at the band edge, with both
 horizontal variables increasing away from the center.
 (a) Bloch wave number $p$ as a function of frequency $\omega$
 on the first Bloch band of the periodically extended lattice. On the one-cell circle, periodic perturbations sample only $p=0$ in the first Brillouin zone.
 (b) Continuation to $\omega=2i\zeta$ on the upper hyperbolic branch.
 The spatial growth rate is
 $|\operatorname{Im}p|=\log r(M)/\pi=S_{\rm ch}/(\pi\kcs)$.
 The solid curves give the exact Floquet result. The dashed lines
 are the tangents at the common edge, both with slope
 $1/v_s\simeq 1.941$ in their respective variables.}
 \label{fig:acoustic-holonomy}
\end{figure}

The same result takes a simple form near the translation mode.
Writing $\mu=e^{ipa}$ and using $na=2\pi$ gives
$\log r(M)=\pi|\operatorname{Im}p|$. Continuing the acoustic
dispersion $p(\omega)=\omega/v_s+O(\omega^3)$ to $\omega=2i\zeta$
therefore yields
\begin{equation}
 S_{\rm ch}=\frac{2\pi^2\kcs}{v_s(n)}\,\Theta_{\rm ch}
 +O(\Theta_{\rm ch}^3).
 \label{eq:entropy-vs}
\end{equation}
Figure~\ref{fig:acoustic-holonomy} shows the two branches meeting at the translation mode.

Appendix~\ref{app:duality} proves the exact relation
\begin{equation}
 \Tr M\!\left(\frac1{4\nu}\right)=(-1)^n\Tr M(\nu),\qquad
 \log r\!\left(M\!\left(\frac1{4\nu}\right)\right)=\log r(M(\nu)).
\label{eq:monodromy-duality}
\end{equation}
The entropy has the same magnitude at dual spectral values, with the same slope with respect to $\Theta_{\rm ch}$ at the paired lower edge. The eigenvalues on the lower hyperbolic branch carry the common sign $(-1)^n$.

Restoring the two chiral copies in Eq. \eqref{eq:entropy-holonomy-definition},
the total holonomy entropy of the winding background is
\begin{equation}
 S=S_{\rm ch}^{+}+S_{\rm ch}^{-}
 =\kcs\,nK(k_n)\sum_{s=\pm}
 \left|\operatorname{Re}Z(\beta_s,k_n)\right| .
 \label{eq:winding-total-entropy}
\end{equation}
For this common-winding family, the winding $n$ fixes the elliptic modulus $k_n$ and the sound velocity of the lattice. The two embeddings $\nu_\pm$ select the points $\beta_\pm$ at which the angular holonomies sample the Floquet dispersion. For the lower branch, Eq. \eqref{eq:monodromy-duality} allows evaluation at $1/(4\nu_-)$. For a common elementary thermal cycle generated by $\partial_t$, the temporal eigenvalue scales must coincide. At common winding and within the vacuum causal parameter range $0<\nu_-<1/2<\nu_+$, the condition $\zeta_+=\zeta_-$ gives $\nu_-=1/(4\nu_+)$. The duality in Eq. \eqref{eq:monodromy-duality} then makes the two contributions equal, so $S=2S_{\rm ch}^{+}$. Along this family, let $\Theta$ denote the common chiral temporal scale. As the embeddings approach the paired band edges at fixed winding, Eq. \eqref{eq:entropy-vs} gives a slope of $4\pi^2\kcs/v_s(n)$ for the total holonomy entropy as a function of $\Theta$. The propagation speed of a displacement along the boundary lattice therefore fixes this slope for the two sectors together.

For the mixed family $(n,0)$, let $h_\pm=\log r(M^\pm)$ and choose a common temporal scale $\zeta=\pi\Theta>0$. The winding copy lies on its upper branch, while the vacuum copy has
\begin{equation}
 \nu_-=\sqrt{\tfrac14+\zeta^2}-\zeta,
 \qquad h_-=\pi\sqrt{1+4\zeta^2}.
 \label{eq:mixed-vacuum-scale}
\end{equation}
The winding exponent is $h_+=S_{\rm ch}/\kcs$ from Eq.~\eqref{eq:holonomy-real}. The BTZ extension established in Appendix~\ref{app:btz} has
\begin{equation}
 S=\kcs(h_++h_-),\qquad
 \ell M_{\rm BTZ}=\frac{\kcs}{4\pi^2}(h_+^2+h_-^2).
 \label{eq:mixed-btz-data}
\end{equation}
Equation~\eqref{eq:mixed-horizon-generator} identifies $\partial_t$ as the laboratory horizon generator, and its smooth Euclidean period is $\pi/\zeta$. In physical laboratory time $\ell t$, its temperature is $T_{\rm lab}=\Theta/\ell$. With respect to the canonical BTZ time $T$, the Hawking temperature $T_H$ obeys
\begin{equation}
 \ell T_H=\frac{h_+h_-}{\pi^2(h_++h_-)},\qquad
 \frac{T_{\rm lab}}{T_H}
 =\frac{\pi^2\Theta(h_++h_-)}{h_+h_-}
 \xrightarrow[\Theta\to0^+]{}\frac{v_s(n)}2.
 \label{eq:mixed-two-temperatures}
\end{equation}
Both temperatures refer to the same horizon, with different normalizations of its generator. 

At fixed $n$ and as $\Theta\to0^+$, the exact spectral result gives
\begin{equation}
 \frac{S}{\kcs}=\pi+\frac{2\pi^2}{v_s(n)}\Theta
                   +2\pi^3\Theta^2+O(\Theta^3).
 \label{eq:mixed-near-edge}
\end{equation}
This expansion varies the embeddings at fixed winding, whereas the charge variation in Eq.~\eqref{eq:cs-laboratory-charge-variation} holds the embeddings fixed. The vacuum supplies the extremal entropy $\pi\kcs$ and the quadratic correction. The entire linear term comes from the lattice. Along this same fixed-winding family, Eqs.~\eqref{eq:mixed-two-temperatures} and \eqref{eq:mixed-near-edge} give $(S-\pi\kcs)/\kcs=\pi^2\ell T_H+O((\ell T_H)^2)$. The sound velocity enters the laboratory slope through the relative normalization of the two horizon generators.

\section{Discussion}

The sound wave probes relative displacements of neighboring solitons, whereas the angular holonomy records transport around the entire circle. The connection between these scales is the Bloch phase per cell. On the acoustic branch its frequency dependence gives the travel time across one cell in the long-wavelength limit. On the hyperbolic branch its continuation gives the exponential growth accumulated over all $n$ cells.

The inverse sound velocity also enters the low-temperature entropy density of the repulsive Lieb--Liniger gas at fixed density, $s_{\rm th}=\pi T/(3v_s)+o(T)$ in the thermodynamic limit \cite{ChenEtAl2019}. This density includes both directions of propagation. Each contributes $\pi T/(6v_s)$ at leading order because the density of modes per unit frequency on each branch is proportional to $1/v_s$. In gravity the derivative appears through the continuation of the acoustic dispersion. For a chiral CFT of circumference $L$ in the extensive regime $LT/v_s\gg1$, the entropy coefficient is $\pi cL/(6v_s)$ \cite{Affleck1986,Cardy1986}. Setting $L=2\pi$ and using the Brown--Henneaux central charge $c=6\kcs$ \cite{BrownHenneaux,Strominger1998} reproduces the gravitational slope. The gravity calculation approaches the band edge at fixed circumference as $\Theta_{\rm ch}\to0$. A statistical interpretation would require an ensemble at fixed boundary sources and an explanation of how these two limits are related. For the mixed family, the horizon interpretation follows from the collar construction of Appendix~\ref{app:btz}, while the inverse-velocity coefficient refers specifically to laboratory temperature; Eq.~\eqref{eq:mixed-two-temperatures} gives its conversion to canonical BTZ temperature.

The acoustic band edge also has a description in terms of a conserved operator. The one-gap Lam\'e fluctuation operator $H$ admits a third-order integral $Q_1$ satisfying $Q_1^2=P_3(H)$, where the roots of the cubic $P_3$ are the three finite band edges \cite{CorreaNietoPlyushchay2007}. Related integrals organize the band-edge states of self-isospectral Lam\'e pairs \cite{CorreaJakubskyNietoPlyushchay2008}. In the present lattice the lowest root corresponds to rigid translation. The conserved operator thus identifies the band edge of the acoustic branch. The slope follows from the Bloch phase accumulated over one lattice period, which fixes the normalization of the quasimomentum near that edge. This suggests asking which conserved spectral data determine the holonomy response for backgrounds with more than one gap.

Do long-wavelength displacements of the lattice survive the gravitational boundary reduction as physical edge modes? The analysis in Ref.  \cite{JoungKim} motivates a concrete test: evaluate the Chern--Simons presymplectic form at fixed boundary sources on a displacement with an allowed nonzero circle momentum and an independent allowed perturbation. To probe the acoustic regime, the wavelength of the displacement must span many lattice cells. A nonzero pairing would establish that the displacement survives the reduction. The entropy slope would then measure the inverse propagation speed of a gravitational edge excitation around the winding background.

The BTZ reference vacuum fixes the normalization of the holonomy entropy. For the mixed $(n,0)$ family, the collar construction and BTZ extension in Appendix~\ref{app:btz} establish a regular exterior and a smooth contractible Euclidean thermal cycle. The sound velocity therefore fixes the leading slope of the horizon entropy with respect to laboratory temperature, above a nonzero limiting entropy supplied by the vacuum copy. For common winding, the boundary metric of aligned profiles loses rank at the angles identified in Sec.~\ref{sec:winding}. A relative shift requires a separate boundary check. For the common-winding family, Eq.~\eqref{eq:entropy-holonomy-definition} gives the entropy defined by the angular spectrum; the BTZ extension constructed here applies to the mixed family.

\acknowledgments

M.C., F.C. and N.S. were supported by Fondecyt Grant No.~1252036. F. C. is also supported by Fondecyt Grant No. 13250014. M.C.\ is grateful to the Department of Physics at New York University for hospitality during the completion of part of this work.

\appendix

\section{BTZ geometry of the vacuum and the mixed sector}
\label{app:btz}

For the stable constant vacuum \eqref{eq:sgvacuum}, Eqs. \eqref{eq:chi-vac}--\eqref{eq:vac-coordinates} give
\begin{equation}
 a^\pm=a_\phi^\pm dx^\pm,
 \qquad
 x^\pm=\phi+\chi_\pm t
 =\Phi\pm\frac{T}{\ell}.
\end{equation}
Because $\nu_+\neq\nu_-$ the transformation from $(t,\phi)$ to $(T,\Phi)$ is invertible, and because $\phi\sim\phi+2\pi$ one has $\Phi\sim\Phi+2\pi$ at fixed $T$.

The two angular connections have eigenvalues $\pm\kappa_\pm$ with \eqref{eq:kappa}.  Any constant hyperbolic element of $\mathfrak{sl}(2,\mathbb R)$ is conjugate to a canonical highest/lowest-weight representative with the same eigenvalues.  We may therefore choose
\begin{equation}
 a^+_{\rm BTZ}=(L_{-1}-\kappa_+^2L_1)dx^+,
 \qquad
 a^-_{\rm BTZ}=(L_1-\kappa_-^2L_{-1})dx^-,
\end{equation}
whose angular holonomies are $\Tr H_\phi^\pm=2\cosh(2\pi\kappa_\pm)$ and coincide with those of the stable Sine--Gordon vacuum.

Introduce a radial coordinate $\hat\rho$ through
\begin{equation}
 b(\hat\rho)=\operatorname{diag}\!\left((\hat\rho/\ell)^{1/2},(\hat\rho/\ell)^{-1/2}\right)=e^{-\log(\hat\rho/\ell)L_0}.
\end{equation}
The metric reconstructed from the two connections is
\begin{align}
 ds^2={}&\frac{\ell^2}{\hat\rho^2}d\hat\rho^2
 -\frac{(\hat\rho^2-\ell^2\kappa_+^2)(\hat\rho^2-\ell^2\kappa_-^2)}
 {\ell^2\hat\rho^2}dT^2
 \nonumber\\
 &+2\ell(\kappa_+^2-\kappa_-^2)dT\,d\Phi
 +\frac{(\hat\rho^2+\ell^2\kappa_+^2)(\hat\rho^2+\ell^2\kappa_-^2)}
 {\hat\rho^2}d\Phi^2.
\label{eq:btzfg}
\end{align}
The areal radius is
\begin{equation}
 r^2=\hat\rho^2+\ell^2(\kappa_+^2+\kappa_-^2)
 +\frac{\ell^4\kappa_+^2\kappa_-^2}{\hat\rho^2}.
\end{equation}
It reaches its minimum at $\hat\rho_H^2=\ell^2\kappa_+\kappa_-$, where $r=r_+$ with $r_+=\ell(\kappa_++\kappa_-)$ and $r_-=\ell(\kappa_--\kappa_+)$. Here $r_-$ is signed and $|r_-|$ is the inner horizon radius.  The metric becomes
\begin{equation}
 ds^2=-N^2dT^2+\frac{dr^2}{N^2}
 +r^2(d\Phi+N^\Phi dT)^2,
\end{equation}
with
\begin{equation}
 N^2=\frac{(r^2-r_+^2)(r^2-r_-^2)}{\ell^2r^2},
 \qquad
 N^\Phi=-\frac{r_+r_-}{\ell r^2},
\end{equation}
which is the standard BTZ geometry \cite{BTZ,BHTZ}.  The entropy and temperature are correspondingly
\begin{equation}
 S_{\rm BTZ}=\frac{2\pi r_+}{4G}
 =2\pi\kcs(\kappa_++\kappa_-),
\end{equation}
\begin{equation}
 T_H=\frac{r_+^2-r_-^2}{2\pi\ell^2r_+}
 =\frac{2\kappa_+\kappa_-}{\pi\ell(\kappa_++\kappa_-)}.
\end{equation}
The radial coordinate $\hat\rho$ of this canonical representative should not be identified with the laboratory radial coordinate used before the bulk gauge/diffeomorphism transformation.  This is why thermal regularity is most robustly characterized by holonomy rather than by zeros of the determinant in a particular radial chart.
The horizon generator is $K_H=\partial_T+\Omega_H\partial_\Phi$, with $\Omega_H=-N^\Phi(r_+)=r_-/(\ell r_+)$. In both copies $\kappa_\pm|\Omega_H\pm1/\ell|=\pi T_H$. After continuation along $K_H$, the integrated Euclidean connection has eigenvalues $\pm i\beta_H\pi T_H$. The period $\beta_H=1/T_H$ therefore gives thermal holonomy $-\mathbf1$, the condition for a smooth Euclidean BTZ geometry \cite{BunsterEtAl2014,IntegrableBH}.

The laboratory generator $\partial_t$ is different. By Eq. \eqref{eq:vac-coordinates} it mixes canonical time translation with rotation, and it is generally not proportional to $K_H$. Its connection $a_t^\pm$ defines the dimensionless scales $\Theta_\pm^{(t)}=|\chi_\pm|\kappa_\pm/\pi$. These scales characterize the temporal auxiliary problem for the chosen laboratory evolution. Even in physical units, $\Theta_\pm^{(t)}/\ell$ need not equal $T_H$. They coincide with the temporal holonomy scale $\Theta_{\rm ch}$ of Sec.~\ref{sec:thermal} evaluated on the vacuum.

\subsection{The laboratory radial chart}

Inserting the vacuum connections into Eqs. \eqref{eq:radial-gauge}--\eqref{eq:dreibein-metric} and using the coordinates $(T,\rho,\Phi)$ gives
\begin{equation}
 ds^2=\frac{\ell^2}{\rho^2}d\rho^2
 +\frac{2\ell^2}{\rho}\,d\rho\,\big(\kappa_+\cos\theta_+\,dx^++\kappa_-\cos\theta_-\,dx^-\big)
 -N^2dT^2+r^2\big(d\Phi+N^\Phi dT\big)^2,
 \label{eq:vac-metric-lab}
\end{equation}
with
\begin{equation}
 r^2(\rho)=\ell^2\big(\kappa_+^2+\kappa_-^2+2\kappa_+\kappa_-\cos\theta_+\cos\theta_-\big)
 +\sigma\,\kappa_+\kappa_-\Big(\rho^2+\frac{\ell^4}{\rho^2}\Big),
 \qquad
 \sigma\equiv-\sin\theta_+\sin\theta_-,
 \label{eq:areal-radius}
\end{equation}
and
\begin{equation}
 N^2=\frac{(r^2-r_+^2)(r^2-r_-^2)}{\ell^2r^2},
 \qquad
 N^\Phi=-\frac{r_+r_-}{\ell r^2},
 \qquad
 r_+=\ell(\kappa_++\kappa_-),
 \qquad
 r_-=\ell(\kappa_--\kappa_+).
\label{eq:rpm}
\end{equation}
The determinant is
\begin{equation}
 \det g=-\frac{\sigma^2\kappa_+^2\kappa_-^2(\rho^4-\ell^4)^2}{\rho^6}
 \qquad\text{in }(T,\rho,\Phi).
\label{eq:det-stable-vac}
\end{equation}
In the laboratory coordinates $(t,\rho,\phi)$ it acquires the factor $[\ell(\chi_+-\chi_-)/2]^2$. Away from $\rho=\ell$ the dreibein is therefore invertible if and only if $\sigma\neq0$ and $\nu_+\neq\nu_-$. If either embedding angle vanishes, the vacuum metric is degenerate for every $\rho$, even though its holonomies are those of BTZ. This is the degeneracy anticipated in Sec.~\ref{sec:embedding-angles}, and it is why the angles $\theta_\pm$ are needed.
 
The reduction to BTZ comes as follows. The $(T,\Phi)$ block of Eq. \eqref{eq:vac-metric-lab} is already the BTZ block at areal radius $r(\rho)$. The cross terms $g_{\rho T}=\ell(\kappa_+\cos\theta_+-\kappa_-\cos\theta_-)/\rho$ and $g_{\rho\Phi}=\ell^2(\kappa_+\cos\theta_++\kappa_-\cos\theta_-)/\rho$ come from the $L_0$ components of $a_\phi^\pm$. They are removed by $T=T_{\rm B}+f(\rho)$ and $\Phi=\Phi_{\rm B}+h(\rho)$, with
\begin{equation}
 f'=\frac{g_{\rho T}-N^\Phi g_{\rho\Phi}}{N^2},
 \qquad
 h'=-N^\Phi f'-\frac{g_{\rho\Phi}}{r^2}.
\end{equation}
In these coordinates the metric becomes
\begin{equation}
 ds^2=-N^2dT_{\rm B}^2+\frac{dr^2}{N^2}+r^2\big(d\Phi_{\rm B}+N^\Phi dT_{\rm B}\big)^2,
 \label{eq:vac-btz}
\end{equation}
which is the BTZ metric \cite{BTZ,BHTZ} with $\Phi_{\rm B}\sim\Phi_{\rm B}+2\pi$. The functions $f$ and $h$ are singular where $N^2=0$, as Schwarzschild-type coordinates are at a horizon. 
 
  The metric \eqref{eq:vac-metric-lab} is invariant under $(\rho,T,\Phi)\to(\ell^2/\rho,-T,-\Phi)$. The region $\rho<\ell$ is therefore the image under simultaneous time and angular reversal of $\rho>\ell$, and we restrict to $\rho>\ell$. As $\rho\to\infty$, $r^2\simeq\sigma\kappa_+\kappa_-\rho^2$, which grows only if $\sigma>0$, that is $c_+c_-<0$. As $\rho$ decreases, $r$ decreases monotonically and reaches its minimum at $\rho=\ell$,
\begin{equation}
 r_{\min}^2=\ell^2\big(\kappa_+^2+\kappa_-^2+2\kappa_+\kappa_-\cos(\theta_++\theta_-)\big),
 \qquad
 r_+^2-r_{\min}^2=2\ell^2\kappa_+\kappa_-\big[1-\cos(\theta_++\theta_-)\big]\geq0,
\end{equation}
with also $r_{\min}^2-r_-^2=2\ell^2\kappa_+\kappa_-[1+\cos(\theta_++\theta_-)]\geq0$. The outer horizon therefore lies at $\rho_H\geq\ell$, where $\rho_H^2+\ell^4/\rho_H^2=2\ell^2(1-\cos\theta_+\cos\theta_-)/\sigma$. For \(\sigma>0\) and \(\theta_++\theta_-\not\equiv0\pmod{2\pi}\), \(\rho_H>\ell\) and the laboratory chart crosses the outer horizon. If \(\theta_++\theta_-\equiv0\pmod{2\pi}\), the radial fold is at \(r_{\min}=r_+\). The chart with \(\rho>\ell\) then ends at the horizon.

In terms of the boundary causal structure, we have that at large $\rho$, Eq. \eqref{eq:vac-metric-lab} gives $ds^2\simeq\sigma\kappa_+\kappa_-\rho^2\,dx^+dx^-$, so the boundary metric is conformal to
\begin{equation}
 g^{(0)}=dx^+dx^-=(d\phi+\chi_+dt)(d\phi+\chi_-dt).
\end{equation}
The angular circle is spacelike when $\sigma>0$. The laboratory generator $\partial_t$ is timelike when $\chi_+\chi_-<0$. Choosing the orientation in which $T$ increases with $t$, the laboratory patch is
\begin{equation}
c_+c_-<0,\qquad \nu_-<\frac12<\nu_+,
\label{eq:physicalpatch}
\end{equation}
where exchanging the copies reverses the orientation of $T$ relative to $t$.\footnote{For fixed masses $m_\pm$, the right-hand side of Eq. \eqref{eq:sg-lab} and the functions $\alpha^\pm,\alpha_u^\pm$ acquire a factor $m_\pm^2$. On the stable vacuum this gives $\kappa_\pm=(4\nu_\pm^2+m_\pm^2)/(8\nu_\pm)$ and $\chi_\pm=(4\nu_\pm^2-m_\pm^2)/(4\nu_\pm^2+m_\pm^2)$, so $\chi_\pm$ changes sign at $\nu_\pm^2=m_\pm^2/4$.} The null directions of $g^{(0)}$ obey $d\phi/dt=-\chi_\pm$. By contrast, the Sine--Gordon equation in the coordinates $(t,\phi)$ has characteristic speed one.

For the mixed family of Sec.~\ref{sec:thermal}, choose $\theta_\pm=\pm\pi/2$.
Equation~\eqref{eq:winding-boundary-frame} gives $2e_{t,\phi}^+=
\nu_++\partial_\phi u_n^+/2\pm\alpha_u^+$, both positive since
$\nu_+>1/2$, $|\alpha_u^+|\leq 1/(4\nu_+)$ and $\partial_\phi u_n^+>0$.
The vacuum gives $e_\phi^-=-\kappa_-<0$ and $e_t^-=\chi_- e_\phi^->0$.
Thus $g_{tt}^{(0)}<0$ and $g_{\phi\phi}^{(0)}>0$ on the entire circle.

On the angular cover let $U_s'= -a^s_\phi U_s$, $U_s(0)=\mathbf1$,
and put $b^2=\operatorname{diag}(q,q^{-1})$, $q=\rho/\ell$ (used here only as a radial variable).
Stationarity and flatness give the developing map
\begin{equation}
 X=e^{t a_t^+(0)}U_+^{-1}b^2U_-e^{-t a_t^-(0)},\qquad
 ds^2=\frac{\ell^2}{2}\operatorname{Tr}[(X^{-1}dX)^2].
\label{app:mixed-collar}
\end{equation}
In temporal eigenbases, chosen so that both temporal matrices are
$\zeta H$, $H=\operatorname{diag}(1,-1)$, write
\begin{equation}
 X=\begin{pmatrix}a&b_1\\c&d\end{pmatrix},\quad
 \alpha=ad=1+b_1c,\quad
 u=\tfrac12\log|a/d|,\quad v=\tfrac12\log|b_1/c|.
\end{equation}
Here $\partial_tX=\zeta[H,X]$, so $u$ and $\alpha$ are independent
of $t$, while $v=2\zeta t+v_0(q,\phi)$. In a component with $\alpha>1$,
\begin{equation}
 ds^2=\ell^2\left\{\frac{d\alpha^2}{4\alpha(\alpha-1)}
             +\alpha\,du^2-(\alpha-1)\,dv^2\right\}.
 \label{eq:mixed-model-metric}
\end{equation}

The rank-one radial decomposition $X(0,q,\phi)=qP(\phi)+q^{-1}Q(\phi)$
gives $\alpha=q^2C(\phi)+O(1)$, uniformly with all needed angular
derivatives on a fundamental interval. Since
$g_{tt}=-4\ell^2\zeta^2(\alpha-1)$, the assumed strict boundary
timelikeness implies $C(\phi)>0$ at every angle. Compactness gives a
positive uniform lower bound. Thus a sufficiently external collar lies
in one exterior wedge, $\partial_q\alpha>0$, and every sufficiently
large value of $\alpha$ determines a unique smooth $q(\alpha,\phi)$.
In particular, the leading diagonal and off-diagonal entries never
vanish, and
\begin{equation}
 u=u_0(\phi)+O(q^{-2}),\qquad
 v=2\zeta t+v_0(\phi)+O(q^{-2}).
\end{equation}
Up to a positive conformal factor, the boundary metric is therefore
\begin{equation}
 du_0^2-(2\zeta\,dt+v_0'\,d\phi)^2.
\end{equation}
Its determinant is $-4\zeta^2(u_0')^2$. Nondegeneracy forces
$u_0'\ne0$ everywhere; continuity then fixes its sign on the entire
circle. 

The angular identification acts by left and right multiplication by
diagonal matrices in these bases, since the angular holonomies commute
with the stationary temporal generators at the base point. Write their
exponents as $pH$ and $mH$, with $|p|=h_+$ and $|m|=h_-$.
The upper branch and the constant vacuum have positive multipliers.
Thus $(\delta u,\delta v)=(p+m,p-m)$. Boundary spacelikeness at fixed
$t$ gives $|u_0'|>|v_0'|$; since $u_0'$ has fixed sign, integration gives
$|\delta u|>|\delta v|$ and hence $pm>0$. In the vacuum copy $a_t^-=\chi_-a_\phi^-$ with $\chi_-<0$, so
$U_-(2\pi)=\exp(-2\pi\zeta H/\chi_-)$ gives $m=h_->0$.
Thus $p=h_+>0$ in the chosen angular orientation. Write $\kappa_s=h_s/(2\pi)$, $\Sigma=\kappa_++\kappa_-$
and $\Delta=\kappa_+-\kappa_-$. One full laboratory turn then gives
$(\delta u,\delta v)=2\pi(\Sigma,\Delta)$ and $u_0'>0$.
The entire lifted angular path stays in one lift of the exterior wedge,
which is simply connected in $(\alpha,u,v)$. Its endpoint is therefore
related to its initial point by the lift preserving that wedge.

The uniform asymptotic expansion implies
$\partial_\phi u|_\alpha=u_0'+O(\alpha^{-1})>0$ for sufficiently
large $\alpha$. Thus, at each such $\alpha$, $u$ is a bijection of
angular covers, with increment $2\pi\Sigma$. The invariant coordinate
$y=v-(\Delta/\Sigma)u$ has $\partial_ty=2\zeta$, and hence takes
every real value exactly once at fixed $(\alpha,u)$. These facts prove
that the collar map is globally one-to-one and onto a complete exterior
BTZ collar, including its lifted angular identification. Explicitly,
\begin{equation}
 \frac{T}{\ell}=\frac{\Sigma v-\Delta u}{4\kappa_+\kappa_-},\qquad
 \Phi=\frac{\Sigma u-\Delta v}{4\kappa_+\kappa_-},\qquad
 r^2=r_-^2+(r_+^2-r_-^2)\alpha,
\end{equation}
where $r_+=\ell\Sigma$ and $r_-=-\ell\Delta$ is signed.
The identification is $(T,\Phi)\sim(T,\Phi+2\pi)$.

The standard BTZ exterior extends this collar to a spacetime
with a regular outer Killing horizon at $\alpha=1$. Its canonical
horizon generator, angular velocity and temperature are
\begin{equation}
 K_H=\partial_T+\frac{r_-}{\ell r_+}\partial_\Phi,
 \qquad T_H=\frac{r_+^2-r_-^2}{2\pi\ell^2r_+}.
\end{equation}
In this equal-scale mixed sector one also has
\begin{equation}
 \partial_t=\frac{\ell\Sigma\zeta}{2\kappa_+\kappa_-}K_H ,
 \label{eq:mixed-horizon-generator}
\end{equation}
so the temperature in the dimensionless laboratory time is
$T_t=\zeta/\pi$. These statements concern the constructed BTZ
extension of the collar; they do not establish that the full original
radial development remains injective or nondegenerate up to $\alpha=1$.

\section{Exact duality and the lower-branch sign}
\label{app:duality}

The two edges in Eq. \eqref{eq:temporal-winding-edges} are exchanged by
$\nu\mapsto\tilde\nu=1/(4\nu)$. Equation \eqref{eq:temporal-spectral-polynomial}
shows that this exchange preserves $\zeta^2$. To see what it does to
the angular multipliers, consider the stationary $+$ connection in
Eq. \eqref{eq:aphi-nu}, for which $q=u'$. We may work before the
constant embedding rotation of Sec.~\ref{sec:embedding-angles}, since
that rotation preserves the monodromy spectrum.

Using the generators of Eq. \eqref{eq:matrix-basis} and the rotation
$R_+$ of Sec.~\ref{sec:embedding-angles}, the same angular connection reads
\begin{equation}
 a_\phi(\nu)=-\nu L_0
 +\tilde\nu\,R_+(u)^{-1}L_0R_+(u)
 +\tfrac14 u'(L_1+L_{-1}).
\end{equation}
The two spectral coefficients thus multiply a fixed generator and its
rotation through the local field angle. To exchange their roles,
first apply $R_+(u)$ to bring the rotated generator back to $L_0$,
then interchange the two components of the auxiliary vector. In the
basis of Eq. \eqref{eq:matrix-basis}, this interchange is
$L_{-1}-L_1$. The combined change of basis is
\begin{equation}
 g(\phi)=(L_{-1}-L_1)R_+(u(\phi)),\qquad
 a_\phi(\tilde\nu)=g a_\phi(\nu)g^{-1}-(\partial_\phi g)g^{-1}.
\end{equation}
Indeed, the interchange reverses both $L_0$ and the rotation generator
$L_1+L_{-1}$ under conjugation. The derivative term contributes
$u'(L_1+L_{-1})/2$, restoring the original coefficient of $u'$.
Here $g$ is a real change of auxiliary basis with determinant $-1$.

The transport equation \eqref{eq:transport} and its initial condition
give $U_{\tilde\nu}(\phi)=g(\phi)U_\nu(\phi)g(0)^{-1}$.
Across one primitive cell $u$ increases by $2\pi$, so the half-angle
in $R_+(u)$ gives $g(\phi+a)=-g(\phi)$. Over the $n$ cells of the
circle the angular monodromy therefore obeys
\begin{equation}
 M(\tilde\nu)=(-1)^n g(0)M(\nu)g(0)^{-1}.
\end{equation}
Taking the trace and the spectral radius gives Eq. \eqref{eq:monodromy-duality}.
For the shifted $-$ profile in Eq. \eqref{eq:winding-solution}, the
unrotated connection is related to the $+$ connection by constant
conjugation with $R_+(\pi)=L_1+L_{-1}$, up to the choice of angular origin.

On the upper branch, Eq. \eqref{eq:at-nu} gives \((a_t)_{11}=\nu/2+\cos u/(8\nu)>0\).
The first component on the \(+\zeta\) eigenline and the second component on the \(-\zeta\) eigenline remain nonzero under transport. Their signs therefore persist over a cell, giving two positive multipliers. Duality multiplies each cell multiplier by \(-1\), hence each lower-branch circle multiplier has sign \((-1)^n\).

Inside the elliptic interval the cell monodromy preserves the two distinct complex conjugate eigenlines of \(a_t(0)\). Its eigenvalues are \(e^{\pm i\vartheta}\), so \(n\vartheta=j\pi\) gives \(M=(-1)^j\mathbf1\). At the self-dual point \(\nu=1/2\) the cell trace vanishes by duality and \(M_{\rm cell}^2=-\mathbf1\). The full-circle monodromy is \((-1)^{n/2}\mathbf1\) for even \(n\) and has trace zero for odd \(n\).

At a simple upper edge, \(\operatorname{Tr}M=2\cosh(\log r)=2+4\pi^2\zeta^2/v_s^2+O(\zeta^4)\). Its derivative with respect to \(\nu\) is nonzero since \(\zeta^2\) has a simple zero. Differentiating \(\det M=1\) at \(M=\pm\mathbf1\) would give zero trace derivative. The edge monodromy is therefore noncentral parabolic, and duality gives the corresponding lower-edge result.

\end{document}